\documentclass[fleqn,10pt]{wlscirep}
\usepackage[utf8]{inputenc}
\usepackage[T1]{fontenc}
\usepackage{mathptmx}
\usepackage{times}
\usepackage{bm} 
\usepackage[normalem]{ulem}
\usepackage{color,soul}
\usepackage{amsmath}

\usepackage{graphicx}
\usepackage[export]{adjustbox}

\title{Evidence for a single-layer van der Waals multiferroic}

\author[1,2,*]{Qian Song}
\author[1,*]{Connor A. Occhialini}
\author[1,*]{Emre Ergeçen}
\author[1,*]{Batyr Ilyas}
\author[3,4]{Danila Amoroso}
\author[5]{Paolo Barone}
\author[6]{Jesse Kapeghian}
\author[7]{Kenji Watanabe}
\author[8]{Takashi Taniguchi}
\author[6]{Antia S. Botana}
\author[3]{Silvia Picozzi}
\author[1]{Nuh Gedik}
\author[1,$\dag$]{Riccardo Comin}
\affil[1]{Department of Physics, Massachusetts Institute of Technology, Cambridge, 02139, Massachusetts, USA.}
\affil[2]{Department of Material Science and Engineering, Massachusetts Institute of Technology, Cambridge, 02139, Massachusetts, USA.}
\affil[3]{Consiglio Nazionale delle Ricerche CNR-SPIN, c/o Universit\`{a} degli Studi `G. D'Annunzio', Chieti, 66100, Italy}
\affil[4]{Nanomat/Q-mat/CESAM,Université de Li\`{e}ge, B-4000 Liège, Belgium}
\affil[5]{Consiglio Nazionale delle Ricerche CNR-SPIN, Area della Ricerca di Tor Vergata, Via del Fosso del Cavaliere 100, 00133 Rome, Italy}
\affil[6]{Department of Physics, Arizona State University, Tempe, 85287, Arizona, USA}
\affil[7]{Research Center for Functional Materials, National Institute for Materials Science, 1-1 Namiki, Tsukuba 305-0044, Japan}
\affil[8]{International Center for Materials Nanoarchitectonics, National Institute for Materials Science,  1-1 Namiki, Tsukuba 305-0044, Japan}

\affil[$\dag$]{e-mail: rcomin@mit.edu}
\affil[*]{These authors contributed equally to this work.}

\begin{document}

\begin{abstract}
Multiferroic materials have garnered wide interest for their exceptional static \cite{Matsukura2015,Kurumaji2013,Tokura2014} and dynamical \cite{Pimenov2006,Rovillain2010,Kibayashi2014} magnetoelectric properties.  In particular, type-II multiferroics exhibit an inversion-symmetry-breaking magnetic order which directly induces a ferroelectric polarization through various mechanisms, such as the spin-current or the inverse Dzyaloshinskii-Moriya effect \cite{Tokura2014, Khomskii2009}.  This intrinsic coupling between the magnetic and dipolar order parameters results in record-strength magnetoelectric effects \cite{Spaldin2019,Tokura2014}. Two-dimensional materials possessing such intrinsic multiferroic properties have been long sought for harnessing magnetoelectric coupling in nanoelectronic devices \cite{Matsukura2015,Huang2018,Jiang2018}. Here, we report the discovery of type-II multiferroic order in a single atomic layer of the transition metal-based van der Waals material NiI$_2$. The multiferroic state of NiI$_2$ is characterized by a proper-screw spin helix with given handedness, which couples to the charge degrees of freedom to produce a chirality-controlled electrical polarization.  We use circular dichroic Raman measurements to directly probe the magneto-chiral ground state and its electromagnon modes originating from dynamic magnetoelectric coupling. Using birefringence and second-harmonic generation measurements, we detect a highly anisotropic electronic state simultaneously breaking three-fold rotational and inversion symmetry, and supporting polar order. The evolution of the optical signatures as a function of temperature and layer number surprisingly reveals an ordered magnetic, polar state that persists down to the ultrathin limit of monolayer NiI$_2$. These observations establish NiI$_2$ and transition metal dihalides as a new platform for studying emergent multiferroic phenomena, chiral magnetic textures and ferroelectricity in the two-dimensional limit.
\end{abstract}


\flushbottom
\maketitle

\thispagestyle{empty}

\section*{Main text}

The recent discovery of intrinsic magnetic order in atomically-thin van der Waals (vdW) materials \cite{Gong2017,Huang2017} has created new opportunities for the study of collective spin phenomena in free-standing two-dimensional (2D) systems and nanoscale devices \cite{Burch2018,Mak2019}. In past years, significant efforts have been made to achieve direct electrical control and manipulation of magnetic properties in 2D \cite{Huang2018,Jiang2018}, but the mechanisms remain elusive. A more promising avenue towards realizing electrical control of 2D magnetism may be found in vdW materials with intrinsic type-II multiferroicity. In type-II multiferroics, the direct coupling between the magnetic and ferroelectric order parameters is enabled by the presence of a spin configuration lacking inversion symmetry \cite{astrov_1960, rado_observation_1961, newnham_magnetoferroelectricity_1978}, resulting in a large and robust magnetoelectric response \cite{kimura_magnetic_2003,  Tokura2014, Spaldin2019}.

Among possible multiferroic vdW materials, several families have been identified \cite{McGuire2017, Mak2019, CuCrP2S6_multiferroic}, most prominently the transition metal dihalides (MH$_2$, M = transition metal, H = halogen). Of particular promise is the magnetic semiconductor NiI$_2$ \cite{Botana2019, Amoroso2020, McGuire2017, Pollini1984, ju2021}, which is host to a rich phase diagram including a type-II multiferroic ground state \cite{Kurumaji2013,Kurumaji2020}. NiI$_2$ crystallizes in the rhombohedral $R\bar{3}m$ structure at room temperature, forming a triangular lattice of Ni$^{2+}$ ions ($3d^8$, $S = 1$) which are stacked along the $c$-axis and held together by weak interlayer bonding (Fig. \ref{fig:fig1}a). The 2D triangular lattice geometrically frustrates the intralayer magnetic exchange interactions that govern the long-range ordering of the local Ni spins \cite{Friedt1976, Amoroso2020, Botana2019}. This leads to a sequence of magnetic phase transitions in NiI$_2$, first to an antiferromagnetic (AFM) state at $T_{N,1} \simeq 75$ K, and then to a proper-screw helimagnetic ground state below $T_{N,2} \simeq 59.5$ K. The latter exhibits long wavelength helical magnetic structure with propagation vector $\mathbf{Q} = (0.138, 0, 1.457)$ reciprocal lattice units (r.l.u.) \cite{Kuindersma1981} (Fig. \ref{fig:fig1}b). The helimagnetic transition is concomitant with the appearance of an in-plane electrical polarization perpendicular to the ordering vector \cite{Kurumaji2013} as well as crystal symmetry lowering from rhombohedral to monoclinic \cite{Kuindersma1981}.

\subsection*{Optical Characterization of Multiferroic Order in Bulk NiI$_2$}

In this study, we investigate the multiferroic states in bulk and few-layer NiI$_2$ crystals\cite{Liu2020}. Due to the complexity of the ground state, which simultaneously breaks mirror, rotational, and inversion symmetries, we employ a suite of complementary optical techniques to track the multiple signatures of the polar and magneto-chiral orders, as a function of temperature and layer number. Optical birefringence is found to originate from a lowering of the lattice symmetry at both the $T_{N,1}$ and $T_{N,2}$ transitions, and is consistent with a breaking of {\bf c}-axis three-fold rotational symmetry (${C}_{3z}$) and a reduction to a single in-plane two-fold symmetry operation ($C_2$) \cite{Kuindersma1981b}. We further use second harmonic generation (SHG) as a probe of inversion-symmetry breaking, which, together with lowering of rotational symmetry to $C_2$, enables to track the presence of polar order down to the single-layer limit. To confirm the presence of magnetochiral order in the helimagnetic phase of NiI$_2$, we perform Raman spectroscopy with circularly polarized light to resolve a coupled spin-lattice electromagnon excitation with strong optical activity as a signature of spin chirality. These experimental findings are supported by first-principles and Monte Carlo simulations, altogether providing robust evidence for the persistence of type-II multiferroic order down to single-layer NiI$_2$.

\begin{figure}[ht]

\centering
\includegraphics{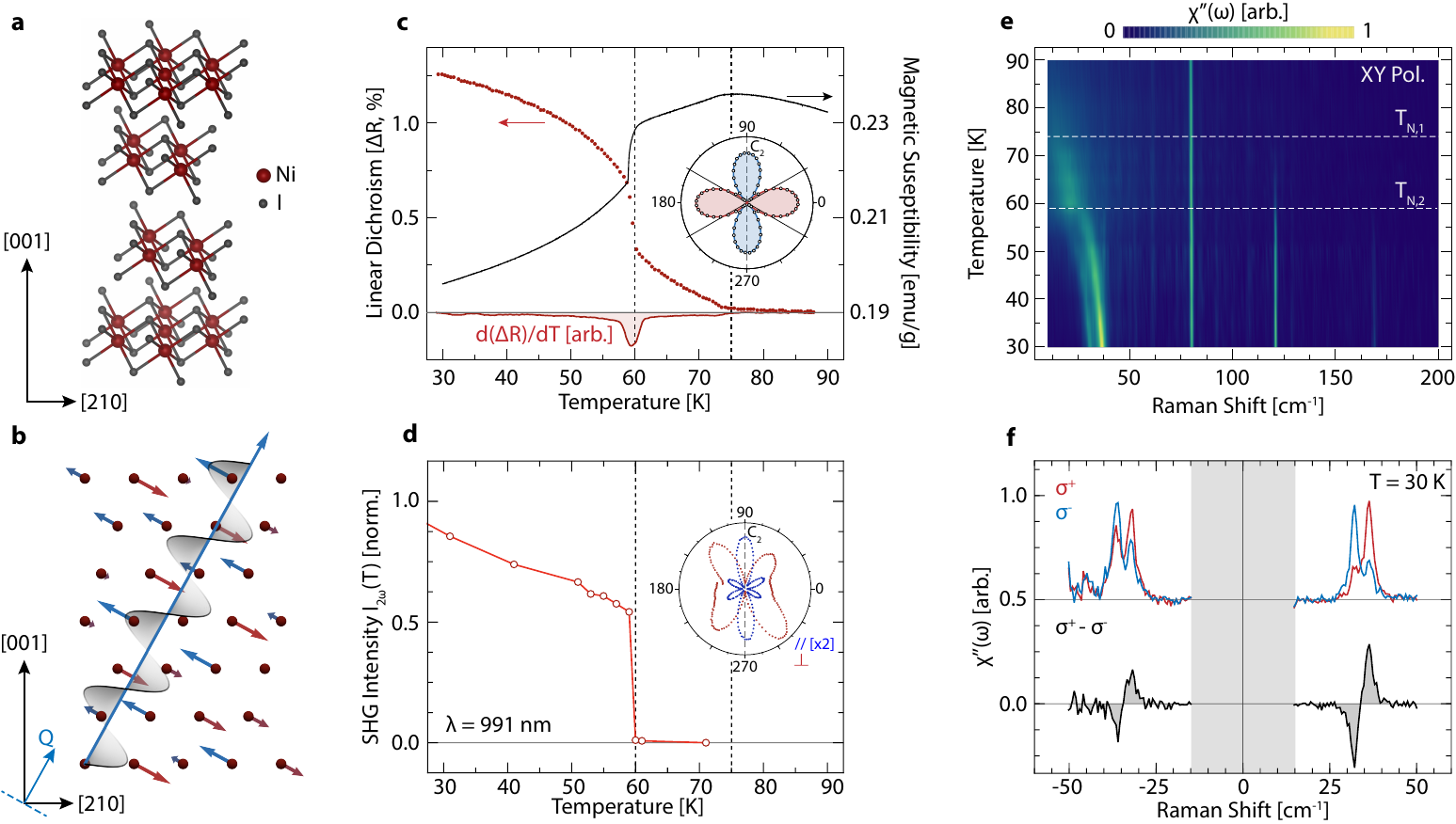}
\caption{\textbf{Crystal structure, magnetic order and optical characterization of bulk NiI$_2$}. \textbf{a}, The high-temperature $R\bar{3}m$ structure of NiI$_2$. \textbf{b}, The helical magnetic structure of bulk NiI$_2$ below $T_{N,2} = 59.5$ K (see text). \textbf{c}, Temperature-dependent linear dichroism ($\Delta R(T)$) measurements on a mono-domain region of bulk NiI$_2$, overlaid on the magnetic susceptibility $\chi_{\text{mag}}$. The inset shows the angular-dependence of $\Delta R$ at $T = 30$ K -- radial lines indicate the crystallographic {\bf a} axes and the dashed line corresponds to the local $C_2$ axis. \textbf{d}, Below band gap ($\lambda = 991$ nm) temperature-dependent SHG in bulk NiI$_2$.  The inset shows rotational anisotropy SHG patterns for a mono-domain region in the $\parallel$ and $\perp$ polarization channels with the dashed line corresponding to the $C_2$ axis orientation. Vertical dashed lines in \textbf{c},\textbf{d} indicate the $T_{N,1}$ and $T_{N,2}$ transitions. \textbf{e}, Temperature-dependent Raman spectra in the XY polarization channel. \textbf{f}, Circular-polarized Stokes and Anti-Stokes Raman spectra of the soft mode excitations for $\sigma^+$/$\sigma^-$ incident polarization (top) and the net Raman optical activity $(\sigma^+ - \sigma^-)$ (bottom).}
\label{fig:fig1}
\end{figure}

We first performed high-resolution, single-domain linear dichroism measurements on bulk NiI$_2$ -- representing the difference in reflectivity $\Delta R$ between two perpendicular, linear polarization states -- across the two transitions at $T_{N,1}$ and $T_{N,2}$ (Fig. \ref{fig:fig1}c). The appearance of optical birefringence indicates the breaking of the three-fold ($C_{3z}$) rotational symmetry in the parent $\bar{3}m$ point group (see Extended Data Fig. \ref{fig:figS2} and Supplemental Information).  The temperature dependence of $\Delta R$ bears a striking resemblance to the magnetic susceptibility ($\chi_{\text{mag}}$, Fig. \ref{fig:fig1}c), with a kink at $T_{N,1}$ and a sharp jump at $T_{N,2}$.  Therefore, the reduction in the lattice rotational symmetry is a direct proxy for the change in the magnetic ground state.  We further record the dependence of the linear dichroism signal on the angle of linear incident polarization with respect to the crystallographic axes, $\Delta R(\theta)$. These results (inset of Fig. \ref{fig:fig1}c and Extended Data Fig. \ref{fig:figS2}) display a maximum dichroism of positive sign (blue lobes) for polarization parallel to the crystallographic {\bf a}-axis and display the expected two-fold rotational symmetry, confirming that the low-temperature phase is characterized by a local, unique $C_2$ axis.

To confirm the polar nature of the low-temperature symmetry group, we measured SHG in bulk NiI$_2$ from a mono-domain region identified through SHG imaging (Fig. \ref{fig:fig1}d and Extended Data Fig. \ref{fig:figS3}). Electric-dipole SHG (ED-SHG) is used as a direct probe of inversion-symmetry breaking \cite{Xiao2018}. Here, the pure electric-dipole SHG mechanism is ensured by tuning the energy of the fundamental beam to a wavelength $\lambda = 991$ nm that is below the optical band-gap and $d$-$d$ transitions \cite{Kuindersma1981b}. This is verified with rotational anisotropy SHG (RA-SHG) measurements, which agree well with the ED-SHG tensor elements predicted for the $C_2$ monoclinic point group in the helimagnetic phase (inset of Fig. \ref{fig:fig1}c, Extended Data Fig. \ref{fig:figS3} and Supplemental Information). For wavelengths in the vicinity of the $d$-$d$ transitions ($\lambda = $ 780 and 826 nm), magnetic-dipole SHG originating purely from spin order, which does not necessitate inversion symmetry breaking, is also present (see Extended Data Fig. \ref{fig:figS3} and Supplemental Information). The combined observation of ED-SHG (lattice inversion-symmetry breaking) and optical birefringence (rotational-symmetry breaking) directly confirms the presence of a polar phase in NiI$_2$ (see Supplemental Information) and an underlying single-{\bf Q}, helical magnetic ground state. The resulting electrical polarization induces a bulk photovoltaic effect which was detected in photocurrent measurements on a thin NiI$_2$ flake (Extended Data Fig. \ref{fig:figS6}).

To underscore the connection between the optical signatures of polar order and the underlying magnetic state, we have performed Raman measurements ($\lambda = 532$ nm) on bulk NiI$_2$. Above $T_{N,1}$, the cross-polarized (XY) Raman spectrum of NiI$_2$ displays a single phonon at $80.2$ cm$^{-1}$ of $E_g$ symmetry \cite{Liu2020}. Our measurements reveal a major change in the Raman response in the magnetically-ordered phase, including the appearance of two new high-energy modes around $120.8$, and $168.8$ cm$^{-1}$ (Fig. \ref{fig:fig1}e) which exhibit polarization selection rules corresponding to single magnon excitations \cite{Cenker2021, Jin2018} (see Extended Data Fig. \ref{fig:figS4}, \ref{fig:figS5}).
At low-energy transfer, a pronounced quasi-elastic signal (QES) develops on approaching $T_{N,1}$\cite{Liu2020} (Fig. \ref{fig:fig1}e). Below $T_{N,2}$, this broad excitation hardens into two sharp and distinct modes (Fig. \ref{fig:fig1}e,f), which display a much more complex set of selection rules compared to the higher energy magnon and phonon excitations (see Extended Data Fig. \ref{fig:figS4}, \ref{fig:figS5}) suggesting they are electromagnons \cite{Pimenov2006,Rovillain2010,Kibayashi2014,Tokura2014,Khomskii2009}. In support of this interpretation, circular polarized excitation reveals a large Raman optical activity (ROA) for the electromagnon peaks (Fig. \ref{fig:fig1}f) that is absent for all other phonon and magnon excitations (Extended Data Fig. \ref{fig:figS4}). The presence of large ROA for the electromagnon underscores it as a direct signature of magneto-chiral order. The reversal of the ROA between Stokes and anti-Stokes scattering (Fig. \ref{fig:fig1}f) is in agreement with the optical selection rules observed for Raman-active magnetic excitations in previous studies \cite{Cenker2021}.

\subsection*{Layer-Dependence of Multiferroicity}
\begin{figure}[ht]

\centering
\includegraphics[max size={\textwidth}{\textheight}]{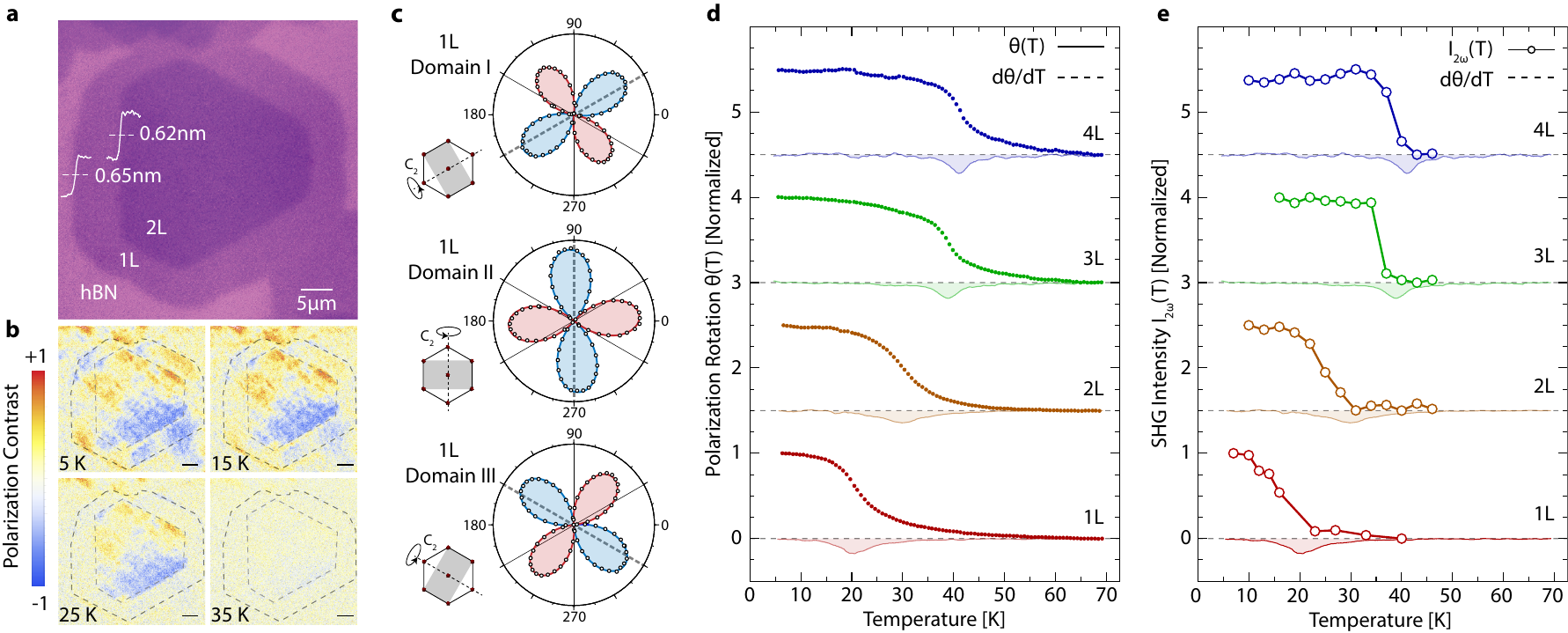}
\caption{\textbf{Birefringence and Second Harmonic Generation in Few- and Single-layer NiI$_2$}. \textbf{a}, Optical image of 1- and 2-layer NiI$_2$ samples grown on hBN with atomic step height profiles measured by AFM (scale bar: $5$ $\mu$m). \textbf{b}, Temperature-dependent polarized microscopy images on the same region in \textbf{a}. Dashed lines demarcate the 1- and 2-layer regions (scale bars: $5$ $\mu$m). \textbf{c}, Angular-dependent linear dichroism $\Delta R(\theta)$ measurements in monolayer NiI$_2$ on three monodomain regions determined from polarized microscopy, as in \textbf{c}.  Radial lines indicate the crystallographic {\bf a}-axes and the dashed lines indicates the orientation of the local $C_2$ axis, as shown in the schematics at the left. \textbf{d}, Temperature-dependent, birefringence-induced polarization rotation $\theta(T)$ measurements acquired in mono-domain regions of 1- to 4-layer NiI$_2$ samples. \textbf{e}, Temperature dependent electric-dipole SHG ($\lambda = 991$ nm) intensity for 1- to 4-layer samples.  Error bars in the SHG data are smaller than the size of the data points. Dashed curves in {\bf d} and {\bf e} represent $d\theta/dT$ calculated from layer-dependent $\theta(T)$ data in {\bf d}. Data are normalized to the value at 5 K and offset vertically for clarity. }
\label{fig:fig2}
\end{figure}

The baseline optical characterization of bulk samples serves as a blueprint for the exploration of the multiferroic state in few-layer NiI$_2$ samples. Figure \ref{fig:fig2}a shows an optical image of 1- and 2-layer NiI$_2$ crystals grown by physical vapor deposition on hBN. The optical anisotropy of NiI$_2$ is directly captured using cross-polarized microscopy (see Methods), signaling the presence of birefringent domains on both the 1- and 2-layer regions at $T = 5$ K (Fig. \ref{fig:fig2}b). As temperature is increased, the birefringent domains vanish from the 1-layer region between $T = 15-25$ K, and from the 2-layer region between $T = 25-35$ K (Fig. \ref{fig:fig2}b). Crucially, a reduction of rotational symmetry from three-fold to two-fold is also observed in single-layer samples, as confirmed by angular dependent linear dichroism $(\Delta R(\theta))$ measurements (Fig. \ref{fig:fig2}c). Angle dependent linear dichroism traces collected in different domains, determined from the cross-polarized images, show the presence of unique two-fold ($C_2$) axes (highlighted by dashed lines in Fig. \ref{fig:fig2}c). The detailed layer dependence of the rotational symmetry breaking transition is obtained through birefringence-induced polarization rotation measurements ($\theta(T)$, see Methods) in Fig. \ref{fig:fig2}d, which reveals a monotonic evolution of the transition temperatures from 1- to 4-layer samples. The transition temperatures are defined as the values of maximum slope of the polarization rotation signal (where $d\theta/dT$ is minimum), yielding $21$ K for single-layer, $30$ K for 2-layer, $39$ K for 3-layer and $41$ K for 4-layer flakes (see Extended Data Fig. \ref{fig:figS10} for 3- and 4-layer optical images and additional $\theta(T)$ data for 1-, 2-, and 4-layer). 

Temperature-dependent ED-SHG ($\lambda = 991$ nm) measurements from 4-, to 1-layer NiI$_2$ samples (Fig. \ref{fig:fig2}e) provide complementary information on the breaking of inversion symmetry down to the single layer limit. For all samples, the SHG intensity displays a clear drop near the transition temperatures derived from the birefringence data. The SHG signal from single-layer samples is genuine as demonstrated by temperature-dependent SHG imaging of multiple monolayer regions (see Extended Data Fig. \ref{fig:figS8}). For these SHG spatial maps, we note a temperature-independent residual SHG contrast from single-layer NiI$_2$ samples which originates at the NiI$_2$/hBN interface due to surface inversion symmetry breaking. The breaking of inversion symmetry demonstrated through SHG and the 3-fold rotational symmetry breaking observed in birefringence measurements (Fig. \ref{fig:fig2}b) unequivocally establish the persistence of a polar ground state in few- and single-layer samples (see Supplemental Information). Concurrently, Raman measurements reveal the persistence of the magnetic soft modes down to the two-layer limit (see Extended Data Fig. \ref{fig:figS9}). The quantitative agreement of the transition temperatures from these independent measurements, as well as the smooth trend of these optical signatures as layer number is reduced from bulk to the monolayer, thus offer strong evidence for the survival of the polar, helimagnetic phase down to the monolayer limit in NiI$_2$.

\subsection*{Theoretical Basis for the multiferroic phase of few-layer NiI$_2$}
The observed reduction of the transition temperature as layer number is decreased is strongly indicative of the relevant role played by the interlayer exchange interaction, as also reported for other 2D magnets~\cite{Huang2017,Deng2018,Jin2018,siv18,antia21,thao21}, while the finite transition temperature detected in the monolayer sample points to a non-negligible magnetic anisotropy. 

\begin{figure}[ht]
\centering
\includegraphics{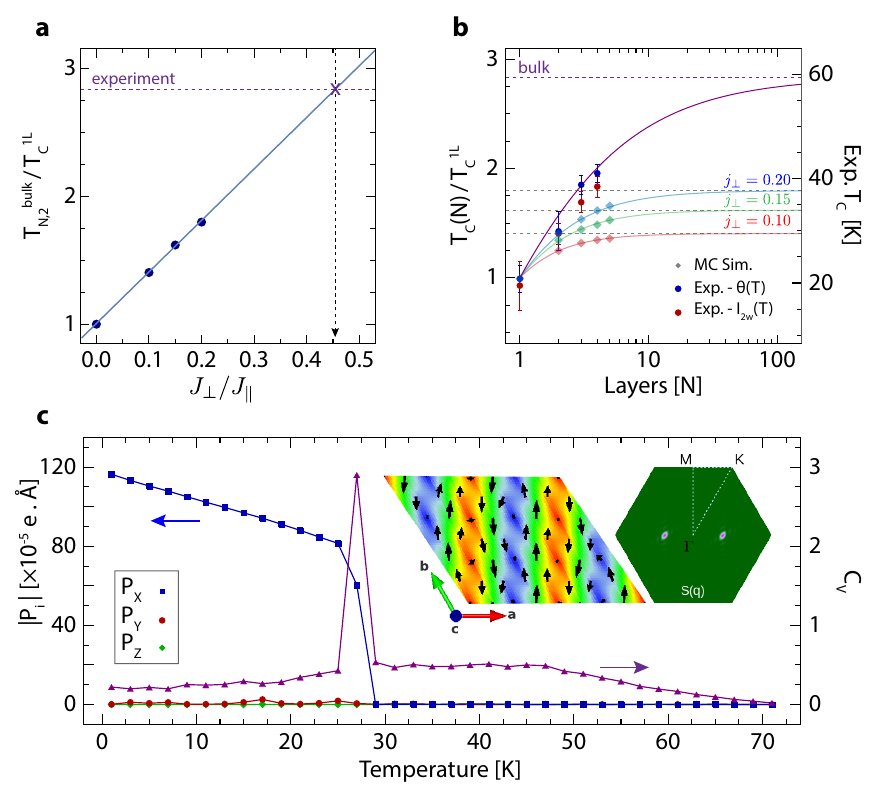}
\caption{\textbf{Layer-Dependent Magnetic Transition Temperatures and the Ground State of Single-Layer NiI$_2$}. 
\textbf{a}, Evolution of the reduced temperature $t$ as a function of the interlayer exchange $J_\perp/J_\parallel$, where $t=T_{N,2}/T_c^{1-layer}$ is the ratio of bulk and 1-layer transition temperatures. Circles denote Monte Carlo estimates, the solid line linearly extrapolates $t$ to larger $J_\perp$. Horizontal dashed line indicates the experimental reduced temperature, $t=\simeq 2.8$, while the vertical one points to the estimated $J_\perp \simeq 0.45 J_\parallel$.
\textbf{b}, Evolution of the critical temperature as a function of the number of layers ($N$). Full symbols (open diamonds) denote the Monte Carlo (experimental) estimates and full lines correspond to the empirical relation provided in Methods (or SI). Dotted horizontal lines represent the bulk transition temperature corresponding to the effective interlayer interaction. \textbf{c}, Electrical polarization components in units of $10^{-5}e$\AA~ (absolute value, closed square) and specific heat (stars) as a function of temperature, relative to the proper-screw spiral order represented in the insets (spin texture -- black arrows represent in-plane components of spins, colormap indicates the out-of-plane spin component, $s_z$=+(-)1 red(blue)- and $S$({\bf Q}), respectively), as obtained via Monte Carlo simulations.}
\label{fig:fig3}
\end{figure}

We performed Monte Carlo (MC) simulations to further investigate the role of the interlayer exchange interaction in affecting the helimagnetic transition temperature. Our starting model is the classical 2D anisotropic Heisenberg model recently introduced for NiI$_2$ monolayer\cite{Amoroso2020}, supplemented by a nearest-neighbour antiferromagnetic interlayer exchange $J_\perp$ in order to describe the bulk phase. The presence of anisotropic interactions guarantees a finite-temperature transition to a long-range ordered magnetic phase even in the monolayer (Fig.~\ref{fig:fig3}c). On the other hand, the transition temperature of the bulk system is found to increase proportionally to $J_\perp$ (Fig.~\ref{fig:fig3}a). Experimentally, the helimagnetic transition temperature of the 1-layer sample is strongly reduced from its bulk value, being $T_{N,2}/T_{c}^{1-layer}\simeq 2.8$. A linear extrapolation of the MC estimates for transition temperatures as a function of $J_\perp$ allows to estimate a quite large interlayer exchange, $J_\perp \sim 0.45~J_\parallel$, where $J_\parallel$ is the dominant ferromagnetic intralayer interaction. Using $J_\parallel\simeq 7~$meV\cite{Amoroso2020}, the estimated interlayer exchange is $J_\perp=3.15~$meV; this is in quite good agreement with the value we directly calculated by means of Density Functional Theory (DFT) calculations for 2-layer NiI$_2$, that is $J_\perp^{(DFT)}\sim 3~$meV. The evolution of transition temperatures as a function of the sample thickness has also been estimated with MC simulations for different number $N$ of layers ($N$ ranging from 1 to 5) and at fixed $J_\perp$. Remarkably, all MC estimates as well as the experimental data are well described by a simple empirical formula that depends only on the ratio of bulk and 1-layer transition temperatures, with no fitting parameters, as shown in Fig.~\ref{fig:fig3}b.

The possible onset of a magnetically induced electric polarization in 1-layer NiI$_2$ has been further investigated in the framework of the generalized spin-current (or Katsura-Nagaosa-Balatsky, gKNB) model~\cite{mike_PRL2011}, where an electric dipole may be induced by spin pairs according to the general expression {\bf P}$_{ij}$ = {\bf M} $\cdot$ {\bf S}$_i$ $\times$ {\bf S}$_j$. The 3$\times$3 {\bf M} tensor has been evaluated from first-principles, and then used to estimate the polarization of NiI$_2$ monolayer from MC simulations. As shown in Fig.~\ref{fig:fig3}d, a proper-screw spiral propagating along the $\bm a$ lattice vector stabilizes below $T_c\lesssim 27~$K, in reasonable agreement with the experimental value of 21~K. The corresponding magnetic point group is the polar $21'$ group, with the twofold rotational axis and the electric polarization coinciding with the spiral propagation vector and hence with the {\bf a} axis, in agreement with the optical measurements. Within the gKNB model, the handedness of the spin spiral uniquely determines if the electric polarization is parallel or antiparallel to the polar axis. We further notice that when the magnetic helix is slanted from the basal plane with a propagation vector {\bf Q} = (0.138,0,1.457), as it occurs in bulk samples, the purely spin-induced polarization as evaluated from our estimated {\bf M}-tensor is always parallel to the crystallographic {\bf a} axis.

\subsection*{Discussion}

NiI$_2$, down to the 2D limit, is therefore an example of improper electronic ferroelectricity, where the electric polarization is driven by the emergence of an inversion-symmetry breaking magnetic order (such as the proper-screw spin helix) in an otherwise centrosymmetric lattice. The propensity for such spin-spiral ordering in the single-layer limit can be associated to the significant frustration of the intralayer exchange interactions on the underlying triangular lattice, along with crucial magnetic anisotropy effects \cite{Amoroso2020}. Furthermore, NiI$_2$ shows superexchange mediated by the extended $5p$ states of the iodine ligands, the latter also introducing a significant spin-orbit coupling (SOC). As a result, non-negligible spin interactions beyond nearest neighbor Ni sites play a relevant role in the stabilization of the resulting proper-screw spin-helix, the latter giving rise to SOC-induced electric polarization, as described by the generalized KNB  model\cite{mike_PRL2011}. In this framework, the observation of SHG and birefringence in monolayer NiI$_2$ is consistent with the proposed theoretical picture, providing a direct indication of a non-centrosymmetric polar magnetic texture that develops below the identified Curie temperature of $21$ K. The observed suppression of the multiferroic transition temperature with reducing layer number further allowed us to clarify the role of the interlayer exchange interactions, which are known to contribute to the orientation of the spin-rotation plane and magnetic ordering vector in the bulk \cite{Kuindersma1981,Friedt1976}. 

Our results point to the crucial role of SOC and orbital extension of the ligands in modulating both the long-range magnetic interactions and the exchange anisotropy, ultimately determining the magnetic ground state among closely lying, competing phases \cite{Amoroso2020}. This suggests that changing the ligand could be a powerful tuning knob to realize new exotic magnetic ground states in 2D, including helices, cycloids and skyrmions with unique topological and multiferroic properties \cite{Amoroso2020,Botana2019,prayitno_controlling_2021}. In addition, our observations of a complex magnetic ground state in an atomically-thin vdW crystal introduces new avenues for exploring multiferroicity which are inherently unique to 2D systems. These include the robust and direct electrical control of magnetism through electrostatic doping or external fields and currents, or the realization of new interfacial multiferroic properties in artificial vdW heterostructures. Moreover, 2D materials with gate-tunable magnetoelectric coupling might offer a new platform for continuous tuning of multiferroic systems toward quantum critical behavior \cite{narayan_multiferroic_2019}. While our optical measurements establish the persistence of multiferroicity in NiI$_2$ to the monolayer limit, further characterization through piezo-response force microscopy or high-sensitivity pyroelectric current measurements will be of great importance to confirm the electrical manifestations of multiferroicity and assess its potential for these future device applications.

\subsection*{Conclusion}

In conclusion, our results represent the first observation of an intrinsic multiferroic phase in the single-layer limit. Multiple complementary optical signatures confirm the persistence of ferroelectricity and inversion-symmetry breaking magnetic order down to monolayer NiI$_2$ with a transition temperature of $T = 21$ K. These results usher the physics of type-II multiferroics into the playground of vdW materials. These findings open a new frontier for studying and optimizing the interplay of magnetic and dielectric properties in next-generation nanoscale devices. 

\bibliography{NiI2_Ref_20210506.bib}

\newpage

\section*{Methods}

\noindent\textbf{Growth and Characterization of Few-Layer NiI$_2$ Crystals}
Bulk-like NiI$_2$ crystals were grown on 300 nm SiO$_2$/Si and few-layer crystals were grown on hexagonal boron nitride (hBN) substrates via physical vapor deposition (PVD) in a horizontal single-zone furnace equipped with a 0.5 in. diameter quartz tube at ambient pressure. hBN was mechanically exfoliated and transferred onto 300 nm SiO$_2$/Si substrates, then annealed in vacuum at 700 °C for 1 h prior to the growth. In a typical synthesis, 0.1 g of NiI$_2$ powder (99.5\%, anhydrous, Alfa Aesar) was positioned at the center of the furnace as the source material and the SiO$_2$/Si substrate was placed downstream at the maximum temperature gradient point. The furnace was purged by pumping the quartz tube below 0.1 Torr and then refilled with 99.99\% Ar gas three times. When growing bulk-like NiI$_2$ crystals, the furnace was heated to 450$^{\circ}C$ in 15 minutes and held at that temperature for 10 minutes. For few-layer NiI$_2$ samples, the furnace was heated to 380-420$^{\circ}C$ in 15 minutes and then the SiO$_2$/Si substrate was taken out immediately and stored inside a nitrogen-filled glove box (O$_2$ < 0.5 ppm, H$_2$O < 0.5 ppm). The sample thickness was determined by atomic force microscopy (AFMWorkshop HR), which was performed inside a separate nitrogen-filled glovebox (O$_2$ < 100 ppm, H$_2$O < 1 ppm), using a silicon probe in tapping mode. The electron diffraction (JEOL HRTEM) was performed on a 7 nm thick PVD grown NiI$_2$ flake using a transmission electron microscope. The flake was picked up and dropped down using the dry transfer technique onto a SiNx membrane with 2 $\mu$m pores (NORCADA)\cite{zomer2014fast}.

\medskip

\noindent\textbf{Growth and Characterization of Single Crystal NiI$_2$}
Single crystal NiI$_2$ was grown by chemical vapor transport (CVT), from elemental precursors with molar ratio Ni:I=1:2, at a temperature gradient 700$^{\circ}$C to 500$^{\circ}$C. The magnetic susceptibility was measured during field cooling at 0.9 T applied out of plane, using a Magnetic Property Measurement System (MPMS-3, Quantum Design Inc.). X-ray diffraction of CVT grown crystals was performed in Bragg geometry using Cu K$_\alpha$ radiation (PANalytical), and the refined unit cell at room temperature is $a = 3.91$ $\textup{\AA}$, $c = 19.93$ $\textup{\AA}$. 

\medskip

\noindent\textbf{Device fabrication for bulk photovoltaic effect measurements.}
The photocurrent device was fabricated by depositing two Ti(5nm)/Au(50 nm) electrodes on a sapphire substrate using a 20 $\mu$m wide wire mask. A PVD-grown bulk-like NiI$_2$ flake was then picked up and dropped down across the gap using a polymer-based dry transfer technique. To minimize the exposure to moisture, the polymer was dissolved in anhydrous chloroform inside the glovebox. To ensure a good and uniform electric contact, a carbon copy of the metal pads was created using the same dry transfer technique and stacked on top of the sample to provide a vertically symmetric contact. We used a 0.3 mW linearly-polarized 532 nm laser for the photocurrent measurement. The current was measured using a Keithley 2401 current meter, and the magnetic field was applied perpendicular to the electric field in plane using a 1 T electromagnet from Montana Instruments.

\medskip

\noindent\textbf{Raman Spectroscopy Measurements}
Polarized Raman experiments were performed in a back-scattering geometry using a confocal microscope spectrometer (Horiba LabRAM HR Evolution) with a 50x objective lens and 532 nm laser excitation at a power of $300 \mu$W ($40 \mu$W) for bulk (few-layer) samples, respectively. Scattered light was dispersed by a 1800 lines/mm grating and detected with a liquid nitrogen cooled charge-coupled device (CCD) camera. The spectrometer integration time was 30/60 minutes for bulk/few-layer samples, and each scan was taken twice and then averaged before analysis. Polarized Raman spectra were recorded with a linearly polarized incident beam. For angle-resolved polarized Raman spectroscopy (ARPRS) measurements, an achromatic half-wave plate was placed just before the objective and rotated in steps of 7.5° from 0° to 180°. An analyzer was placed in front of the spectrometer entrance and kept vertical/horizontal for parallel (XX)/perpendicular (XY) configurations, respectively. For circularly polarized measurements, an achromatic quarter-wave plate was placed in front of the objective with fast axis oriented at +/- 45 degrees with respect to the incident linear polarization for $\sigma^+$/$\sigma^-$ circular incident polarization, respectively.  For the reported circularly polarized spectra, no analyzing polarizer is used. Temperature-dependent measurements in the range $5 - 300$\,K were performed using a Montana Instruments closed-cycle optical cryostat.

\medskip

\noindent\textbf{Birefringence Measurements} A supercontinuum light source (NKT Photonics, Fianium) monochromatized to $\lambda = 532$ nm/$550$ nm and a bandwidth of approximately $1$ nm was used as excitation for angular-dependent linear dichroism and birefringence-induced polarization rotation measurements, respectively. All measurements were performed at normal incidence in a Montana Instruments closed-cycle optical cryostat. Linear dichroism measurements were performed with a photo-elastic modulator (PEM-100, Hinds Instruments) on the incident path of the optical setup. The beam incident on the PEM is prepared in linear polarization making an angle of $45^\circ$ with the respect to the PEM fast axis and amplitude modulated with a mechanical chopper. The PEM retardance was set to 0.5$\lambda$ to modulate the incident polarization between $\pm 45^\circ$ linear polarization states. The light is then focused onto the sample using a 50x objective lens. The backscattered light is measured by an amplified photodiode (ThorLabs PDA100A2), whose output is connected to a lock-in amplifier (Stanford Instruments SR865A) referenced to the second harmonic of the fundamental PEM frequency $f = 50$ kHz. The total reflectance of the sample, used as a normalization, is monitored by a second lock-in amplifier referenced to the chopping frequency $f = 557$ Hz.  

\noindent To perform angular-dependent linear dichroism measurements, the angle of the perpendicular linear polarization states created by the PEM is varied across the crystal using a zero-order half-wave plate placed just before the objective.  In order to ensure the angular-dependence is recorded from a uniform, mono-domain region of the sample, polarized microscopy images were first recorded at the base temperature $T = 5$ K.  The sample was held at this temperature for the duration of the angular-dependent measurements in order to maintain the same distribution of birefringent domains. Birefringence-induced polarization rotation measurements on the few- and single-layer samples were performed with the PEM on the detection path of the optical setup with a retardance amplitude of 0.486$\lambda$, using a lock-in amplifier at the second harmonic of the fundamental PEM frequency for signal collection. The unrotated incident beam is kept parallel to the PEM fast-axis and analyzed by a polarizer at 45 deg. before being measured by the photodiode. This setup provided a sensitivity down to 10 $\mu$rad.

\noindent Polarized microscopy was performed with a broadband visible LED light source, a standard CMOS-based monochrome camera and Glan-Thompson polarizers on both the input and output light paths in reflection geometry. A detuning of 0.5 (2.0) degrees from a cross-polarized configuration was used to maximize the contrast from birefringent domains and a 5 (2) second integration time was used for the images of few-layer (bulk) samples (Extended Data Fig. \ref{fig:figS7}). Cross-polarized contrast images were obtained as a function of temperature by auto-correlating the images for different temperatures to overlap with a high-temperature ($T > T_{N,1}$) reference, and subtracting the low and high temperature images. 

\medskip

\noindent\textbf{Layer-Dependent Second Harmonic Generation Microscopy.}
In second harmonic generation (SHG) microscopy, an objective lens (Olympus LMPlanFL-N 50x) focuses an ultrashort laser beam onto the sample located in a cryostat (Janis ST-500). For wavelength dependent SHG experiments, we used a optical parametric amplifier (ORPHEUS, Light Conversion) seeded with a regenerative amplifier (PHAROS SP-10-600-PP, Light Conversion) These allowed us to tune the output wavelength in a wide spectral range. The estimated spot size on the sample is $\sim 2 \mu m$. The laser fluence incident on the sample was set to 1 mJ/cm$^2$. No sample damage or degradation was observed during the measurements. Upon reflection, the second harmonic component of the beam radiated from sample was selected out by a dichroic mirror and a monochromator with 2 nm spectral resolution. The second harmonic photons were detected and counted using a photomultiplier tube (Hamamatsu PMT) and a dual-channel gated photon-counter (Stanford Research SR400). To decrease the background and dark counts, the photon counter was synchronized and gated with laser pulses. To perform SHG imaging (Extended Data Fig. \ref{fig:figS8}), we keep the sample location fixed and rotate a motorized mirror to scan the laser across the sample. The polarization angle of the pulses were controlled using a wire-grid polarizer and a half-waveplate respectively to obtain the polarization resolved second harmonic traces (Extended Data Fig. \ref{fig:figS7}).

\medskip

\noindent\textbf{Spin model and Monte Carlo simulations.}
To describe the magnetic properties of the system we adopted the 2D anisotropic Heisenberg model derived for NiI$_2$ monolayer\cite{Amoroso2020} supplemented by a nearest-neighbour antiferromagnetic interlayer interaction accounting for the exchange between rhombohedral-stacked layers:
\begin{equation}
H = \frac{1}{2}\sum_{ij} \left(\mbox{\bf S}_i^{(e)}\cdot\mbox{\bf J}_{ij}\cdot\mbox{\bf S}_j^{(e)} + \mbox{\bf S}_i^{(o)}\cdot\mbox{\bf J}_{ij}\cdot\mbox{\bf S}_j^{(o)}\right) +
    \sum_i\mbox{\bf S}_i\cdot\mbox{\bf A}_{ii}\cdot\mbox{\bf S}_i+ J_\perp\sum_{<ij>}\mbox{\bf S}_i^{(e)}\cdot\mbox{\bf S}_j^{(o)}
\end{equation}
Here $e$ and $o$ label respectively even and odd layers perpendicular to the {\bf c} axis, the first sum extends to third nearest neighbour within each plane while the last sum is restricted to nearest-neighbours belonging to different layers. We used the interlayer exchange {\bf J}$_{ij}$ and single-site anisotropy {\bf A}$_{ij}$ tensors evaluated from first principles for NiI$_2$ monolayer\cite{Amoroso2020}; the non-negligible anisotropic terms guarantee a finite-temperature long-range magnetic order, with a triple-{\bf Q} topological phase competing with single-{\bf Q} spiral configuration. The anisotropic part of the exchange tensor has been accordingly rescaled to 60\% of the ab initio estimate, thus favouring the spin-spiral solution, in agreement with the breaking of the three-fold rotational symmetry experimentally detected. We set the energy scale as $J_\parallel=-J^{1iso}$, the ferromagnetic nearest-neighbour interlayer interaction. 

\noindent We performed Monte Carlo simulations using a standard Metropolis algorithm on rhombohedral-stacked triangular lattices (rhombohedral supercells in hexagonal setting) in slab geometry to simulate multilayer (bulk) NiI$_2$. We used 10$^5$ MC steps for thermalization and 5$\times$10$^5$ MC steps for statistical averaging at each simulated temperature. Simulations have been performed on 24$\times$24 lattices in slab geometry comprising up to 5 layers of triangular NiI$_2$ (and up to 2880 spins) with two-dimensional periodic boundary conditions (PBC), and on 8$\times$8$\times$8 hexagonal supercells (comprising 1536 spins) with three-dimensional PBC for bulk rhombohedral system. The transition temperature is identified by the peak in the specific heat of the spin model. All MC results for multilayer slabs at different values of interlayer exchange $J_\perp$ are well described by a simple empirical function:
\begin{equation}
\label{eq:empirical}
T_c(N) = T_c^{1-layer} \,\tanh{\left(b\,\ln{N} +\frac{1}{2}\ln{\frac{t+1}{t-1}} \right)},
\end{equation}
where $t=T_{N,2}/T_{c}^{1-layer}$ is the reduced temperature, $N$ is the number of layers and the coefficient $b=1.05$ has been obtained by fitting only the $J_\perp = 0.1~J_\parallel$ data points.

\medskip

\noindent\textbf{Generalised spin-current model for magnetically induced polarization.}
Within the generalised spin-current model gKNB~\cite{mike_PRL2011}, the total polarization is given by:
\begin{equation}
\mbox{\bf P} = \frac{1}{2N}\sum_{ij} \mbox{\bf M}_{ij}\,\cdot\mbox{\bf S}_i\times\mbox{\bf S}_j,
\end{equation}
where $N$ is the number of magnetic sites, and we restricted the sum to interlayer nearest-neighbours.
The {M}-tensor has been evaluated from first principles for a spin pair parallel to the {\bf a} axis using the four-state method\cite{mike_PRL2011} (the tensor of equivalent bonds is readily obtained by enforcing crystalline symmetries). The dominant tensor components found are $M_{22}=348 \times 10^{-5}~$e\AA~ and $M_{23}=-520 \times 10^{-5}~$e\AA, the other components being zero or smaller than $\sim 30~$e\AA. Assuming a proper-screw spiral with positive/negative handedness $\tau=\pm 1$ propagating along the {\bf a} ($x$) axis with pitch $2\delta$, the gKNB model predicts $P_x \equiv P_\parallel = 3\, M_{22}\sin{(\tau\delta)}/2$, in agreement with numerical MC calculations. For the proper-screw spiral propagating along the {\bf Q} = (0.138,0,1.457) direction, with the spins rotating in a plane making an angle $\theta=55^\circ$ as observed in bulk NiI$_2$\cite{Kuindersma1981}, the electric polarization predicted by the gKNB model lies in the $ab$ plane and is perpendicular to the in-plane projection of {\bf Q}, $P_\perp = \sqrt{3}[M_{22}\cos\theta-2 M_{23}\sin\theta]\sin{(\tau 0.138)/2}$. In both cases, handedness $\tau$ determines the sign of {\bf P}, that is always parallel to the crystallographic axis {\bf a}.

\medskip

\noindent\textbf{First-principles calculations.}
We used Density Functional Theory (DFT) to estimate the interlayer exchange interaction and the {\bf M}-tensor of the gKNB model. The interlayer coupling $J_\perp$ has been estimated from the energy difference between a ferromagnetic (FM) and antiferromagnetic (AFM) stacking of ferromagnetically ordered NiI$_2$ layers, ($\Delta$E = E$_{FM}$- E$_{AFM}$). A rhombohedral-stacked bilayer has been constructed starting from the optimised NiI$_2$ monolayer, with lattice parameter $a$ $\sim$ 3.96~\AA~ and an interlayer distance of about 6.54~\AA, in agreement with the previously reported bulk value~\cite{McGuire2017}. A vacuum distance of about 18.45~\AA~ was introduced between periodic copies of the free-standing bilayer along the {\bf c} axis. To check the consistency of our results, we performed DFT calculations using both the projector-augmented  wave (PAW) method as implemented in the VASP code~\cite{vasp1,vasp2} and the all-electron, full potential code WIEN2k~\cite{wien2k_1}, based on the augmented plane wave plus local orbital (APW+lo) basis set. Ni $3p$, $3d$ and $4s$, and I $5s$ and $5p$ have been treated as valence states in VASP calculations, with a plane-wave cutoff of 500 eV and a $18\times18\times2$ $k$-points mesh for Brillouin-zone (BZ) integration. 
In the WIEN2k calculations, a muffin-tin radius of 2.5~\AA~ was used for both Ni and I atoms, as well as an RK$_{max}$ of 7.0. A k-mesh of $20\times20\times2$ was used for the BZ sampling. 
For consistency with previous monolayer calculations~\cite{Amoroso2020}, we used the Perdew-Burke-Erzenhof (PBE) version of the generalized gradient approximation as the exchange-correlation functional~\cite{PBE} and further performed PBE+U calculations~\cite{Rohrbach_2003} employing $U=1.8$~eV and $J=0.8$~eV on the localized Ni-$3d$ orbitals within the Liechtenstein approach~\cite{Liechtenstein_1995}. 
The interlayer exchange defined as  $J_\perp=\frac{\Delta E}{6S^{(1)}S^{(2)}}$ is $\simeq +3.1$~meV within VASP and $\simeq +2.8$~meV within WIEN2k. Similar results were also obtained by introducing the spin-orbit coupling (SOC) and by employing PBEsol~\cite{PBEsol} and optB86~\cite{optB86b} exchange-correlation functionals, that proves robustness of our results. A similar procedure applied to NiBr$_2$ bilayer yields a much smaller interlayer AFM coupling of about $\simeq +1.2$~meV, suggesting a non-trivial role of extended I-5$p$ orbital states in mediating spin exchange across layers of NiI$_2$.

\noindent The {\bf M}-tensor has been evaluated following the no-substitution four-state method\cite{mike_PRL2011} performing DFT+U+SOC calculations with VASP on a $5\times4\times1$ supercell of NiI$_2$ monolayer. We selected a pair of spins ({\bf S$_1$} and {\bf S$_2$}) along the $x$-axis (parallel to the $\bm a$ lattice vector) and calculated the Berry-phase polarization arising from different sets of four noncollinear spin configurations, defining the {\bf M}-tensor, in units of $10^{-5}$~e\AA, as
\begin{equation}
{\bf M} = \begin{bmatrix}
(\bm P_{12}^{yz})_x & (\bm P_{12}^{zx})_x & (\bm P_{12}^{xy})_x \\
(\bm P_{12}^{yz})_y & (\bm P_{12}^{zx})_y & (\bm P_{12}^{xy})_y \\ 
(\bm P_{12}^{yz})_z & (\bm P_{12}^{zx})_z & (\bm P_{12}^{xy})_z
\end{bmatrix}
\quad = \quad
\begin{pmatrix}
20 & 0 & 32 \\
0 & 348 & -520 \\ 
0 & 25 & 0
\end{pmatrix}
\end{equation}
The accuracy of polarization values has been checked by repeating calculations with a larger vacuum of about 32 \AA~ and including dipole corrections.

\flushbottom

\setcounter{figure}{0}

\newpage

\noindent\textbf{Acknowledgements}
\smallskip

\noindent This work was supported by the STC Center for Integrated Quantum Materials, NSF Grant No. DMR-1231319. E.E, B.I. and N.G acknowledge support from the US Department of Energy, BES DMSE (data taking and analysis) and Gordon and Betty Moore Foundation’s EPiQS Initiative grant GBMF9459 (instrumentation). J.K and A.S.B acknowledge NSF Grant No. DMR-1904716 and the ASU research computing center for HPC resources. D.A. and S.P. acknowledge support by the Nanoscience Foundries and Fine Analysis (NFFA-MIUR Italy) project. P.B. and S.P. acknowledge financial support from the Italian Ministry for Research and Education through PRIN-2017 projects “Tuning and understanding Quantum phases in 2D materials—Quantum 2D” (IT-MIUR Grant No. 2017Z8TS5B) and “TWEET: Towards Ferroelectricity in two dimensions” (IT-MIUR Grant No. 2017YCTB59), respectively. D.A., P.B. and S.P. also acknowledge high-performance computing (HPC) systems operated by CINECA (IsC72-2DFmF, IsC80-Em2DvdWd, IsC88-FeCoSMO, and IsB21-IRVISH projects) and computing resources at the Pharmacy Dept., Univ. Chieti-Pescara, and thank Loriano Storchi for his technical support. 

\medskip

\noindent\textbf{Author contributions}
\smallskip

\noindent Q.S. and R.C. conceived the project. Q.S. synthesized the NiI$_2$ samples. Q.S. and C.A.O. performed Raman and birefringence measurements supervised by R.C.. E.E. and B.I. performed SHG measurements supervised by N.G. T.T. and K.W. provided and characterized bulk hBN crystals. D.A., A.S.B. and J.K. performed first-principles calculations. P.B. and D.A. performed Monte Carlo simulations, and discussed the results  with S.P. All authors contributed to the writing of the manuscript. The authors declare no competing financial interests.
 
\medskip

\noindent\textbf{Additional Information}
\smallskip

\noindent Supplementary Information is available for this paper. Correspondence and requests for materials should be addressed to R.C.(rcomin@mit.edu). Reprints and permissions information is available at www.nature.com/reprints

\medskip

\noindent\textbf{Data Availability}
\smallskip

\noindent The datasets generated during and/or analysed during the current study are available from the corresponding author on reasonable request.

\newpage

\begin{figure}[hbt!]
    \centering
    \renewcommand{\figurename}{Extended Data Figure}
    \includegraphics{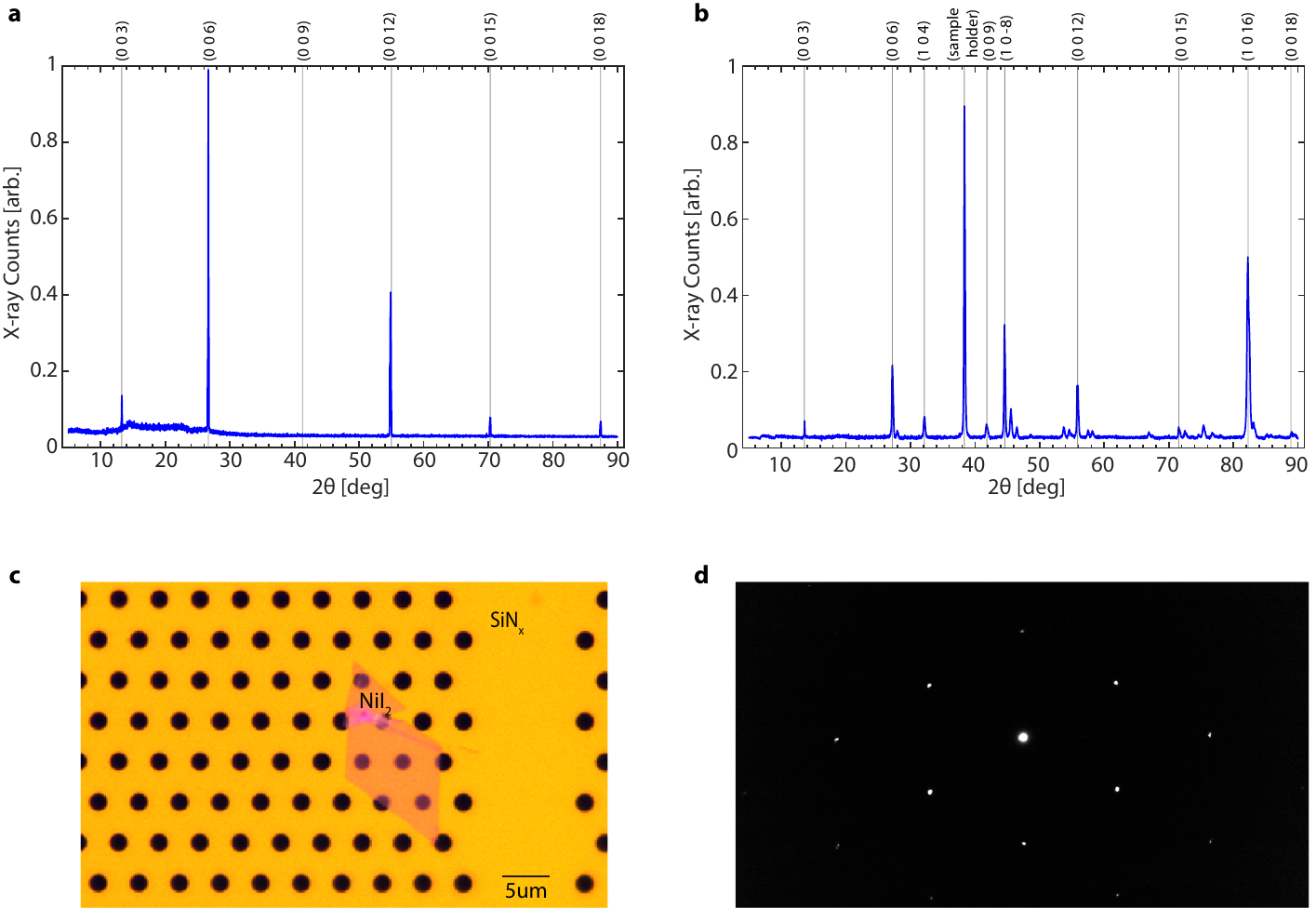}
    \caption{\textbf{X-ray and electron diffraction of NiI$_2$ crystals}. \textbf{a}, X-ray diffraction of a CVT-grown NiI$_2$ single crystal along the {\bf c}-axis. \textbf{b}, X-ray powder diffraction of CVT-grown NiI$_2$. \textbf{c}, An optical image of a 7 nm PVD grown NiI$_2$ flake transferred onto a SiN$_x$ membrane. \textbf{d}, The electron diffraction pattern of the PVD grown NiI$_2$ flake shown in \textbf{c}, using a transmission electron microscope.}
    \label{fig:figS1}
\end{figure}

\newpage

\begin{figure}[hbt!]
    \centering
    \renewcommand{\figurename}{Extended Data Figure}
    \includegraphics{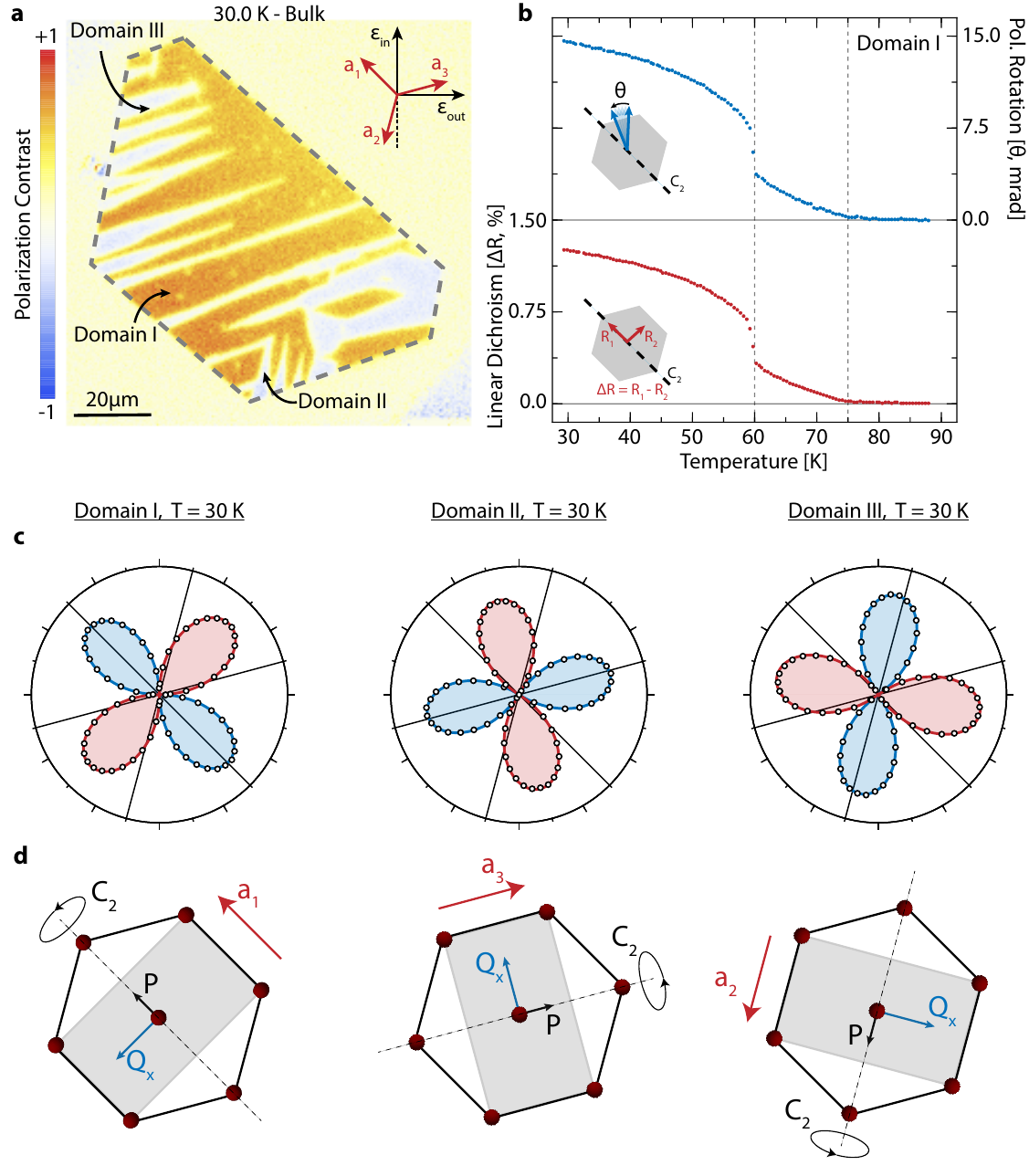}
    \caption{\textbf{Linear dichroism and birefringence-induced polarization rotation measurements in bulk NiI$_2$}. \textbf{a}, Polarized microscopy image of bulk NiI$_2$. The positions where the optical measurements were performed are labelled as Domain I-III. \textbf{b}, Comparison of the temperature-dependent birefringence-induced polarization rotation (top) and linear dichroism (bottom) on Domain I. \textbf{c}, Angular-dependent linear dichroism measurements from the three distinct domains identified in \textbf{a}. Radial lines indicate the crystallographic $a$-axes determined from the edges of the as-grown PVD sample. \textbf{d}, Schematics of the domains as identified from AD-LD measurements in \textbf{c}, denoting the local $C_2$ axis orientation, the polar vector $P$ and the in-plane component of the helimagnetic ordering vector $\mathbf{Q}$.}
    \label{fig:figS2}
\end{figure}

\newpage

\begin{figure}[hbt!]
    \centering
    \renewcommand{\figurename}{Extended Data Figure}
    \includegraphics{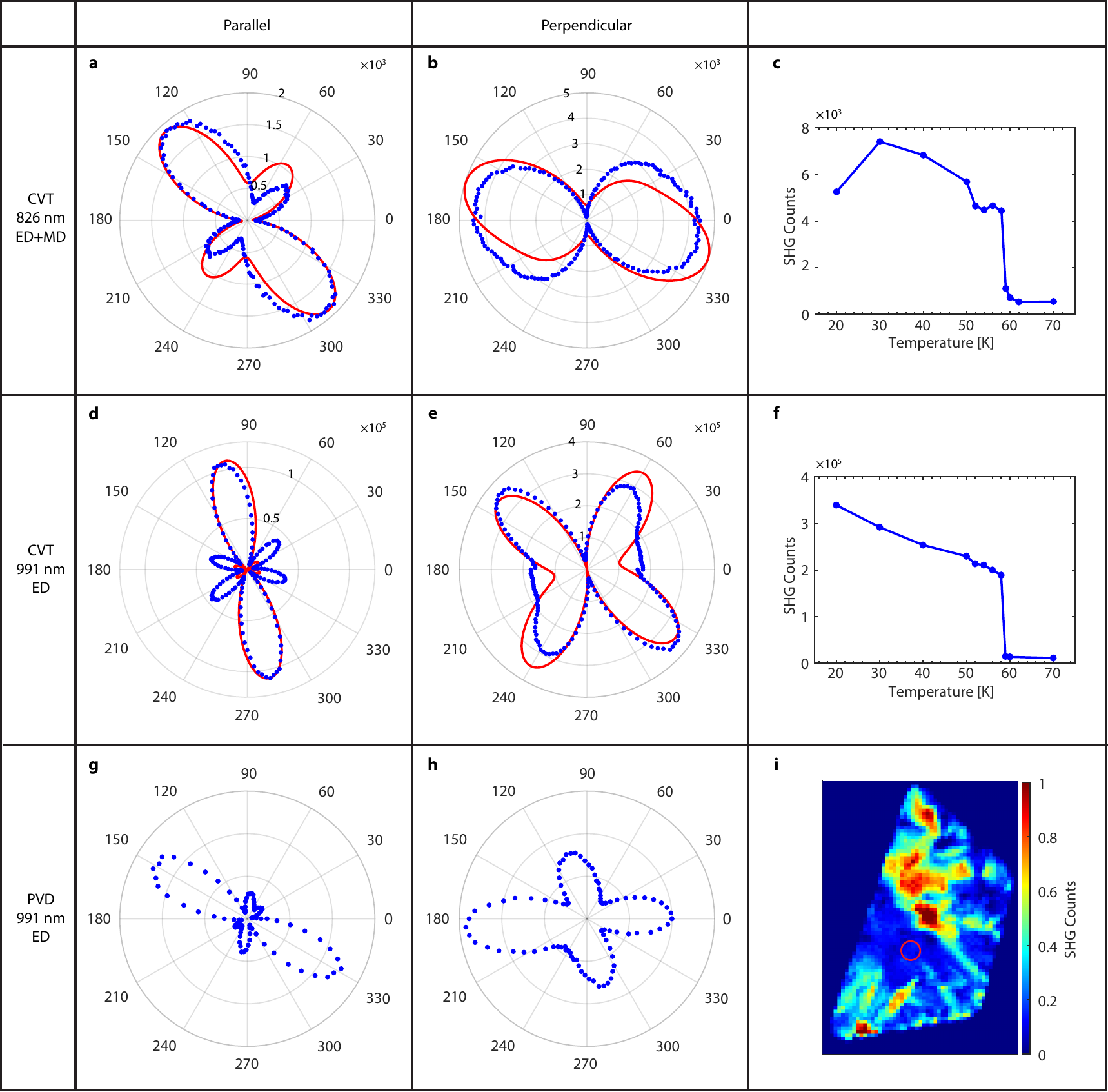}
    \caption{\textbf{Wavelength-dependent Second Harmonic Generation of NiI$_2$}. Rotational anisotropy SHG (RA-SHG), fits to nonlinear tensor elements and their temperature dependence on a single domain CVT grown bulk NiI$_2$ using \textbf{a-c}, 826 nm laser, and \textbf{d-f}, 991 nm laser. The RA-SHG traces obtained with 826 nm can only be fit with a combination of electric dipole (ED) and magnetic dipole (MD) radiation, whereas the RA-SHG traces obtained with 991 nm only exhibit ED component. \textbf{g, h}, RA-SHG on PVD grown bulk NiI$_2$ samples shows the same signatures as the CVT grown samples. \textbf{i} SHG imaging of the PVD sample at 15 K. The red circle shows the single domain region where the RA-SHG was taken.}
    \label{fig:figS3}
\end{figure}

\newpage

\begin{figure}[hbt!]
    \centering
    \renewcommand{\figurename}{Extended Data Figure}
    \includegraphics{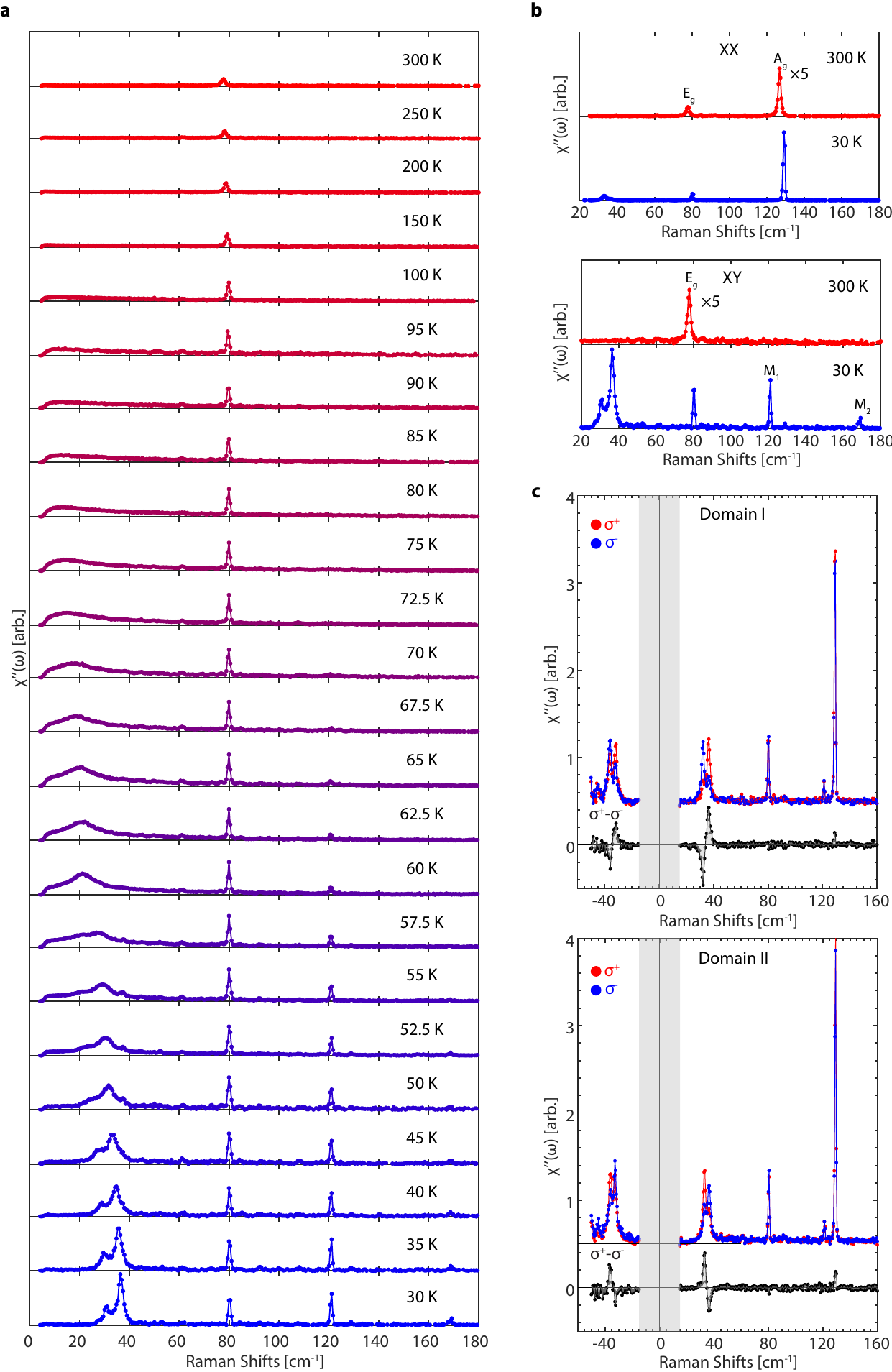}
    \caption{\textbf{Temperature-dependent polarized Raman spectra of bulk NiI$_2$}. \textbf{a}, Raman data in the cross-polarized XY channel from 30 K to 300 K. \textbf{b}, Comparison of the cross-polarized (XY) and parallel-polarized (XX) channels at high and low temperature. \textbf{c}, Circularly polarized Raman spectra at 30 K on domain I and domain II regions for $\sigma^+$/$\sigma^-$ incident polarization (top) and the net ROA $(\sigma^+ - \sigma^-)$ (bottom).}
    \label{fig:figS4}
\end{figure}

\newpage

\begin{figure}[hbt!]
    \centering
    \renewcommand{\figurename}{Extended Data Figure}
    \includegraphics{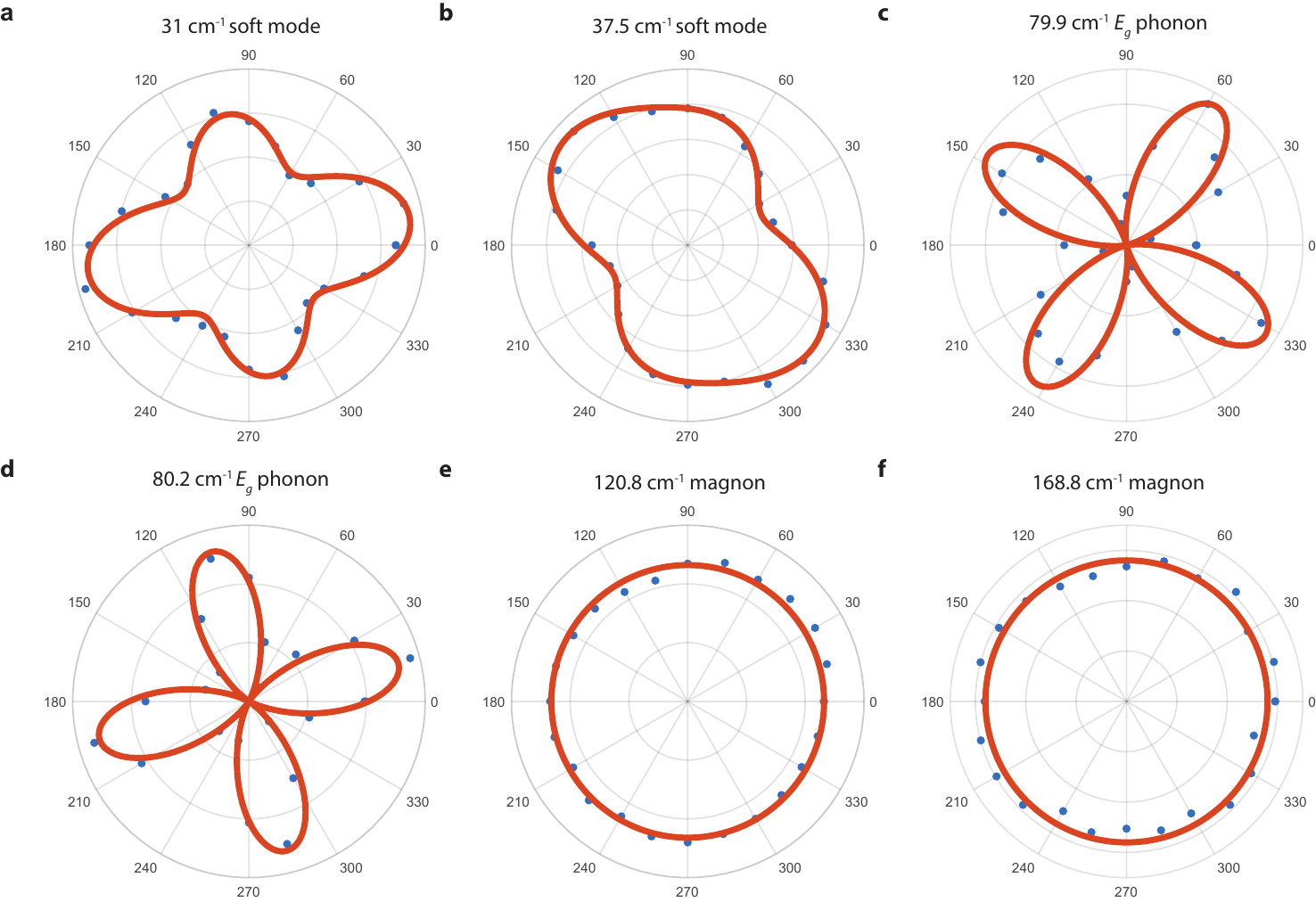}
    \caption{\textbf{Angular Resolved Polarized Raman Spectroscopy (ARPRS) in cross-polarized (XY) configuration}. \textbf{a, b}, The ARPRS polar plots of the 31 cm$^{-1}$ and 37 cm$^{-1}$ modes appearing in the multiferroic phase. Neither agrees with a pure phonon or magnon mode, suggesting they may be consistent with electromagnons. \textbf{c, d}, The 80 cm$^{-1}$ peak is composed of two phonons below $T_{N,2}$, one at 79.9 cm$^{-1}$ and the other at 80.2 cm$^{-1}$. These closely-spaced phonon modes display out-of-phase modulation with respect to the incident linear polarization and both display an $E_g$ symmetry with respect to the high-temperature $R\bar{3}m$ phase. \textbf{e, f}, The 120.8 cm$^{-1}$ and 168.8 cm$^{-1}$ are magnon modes. Red lines: ARPRS fits to the Raman tensors for different mode symmetries.}
    \label{fig:figS5}
\end{figure}

\newpage

\begin{figure}[hbt!]
    \centering
    \renewcommand{\figurename}{Extended Data Figure}
    \includegraphics{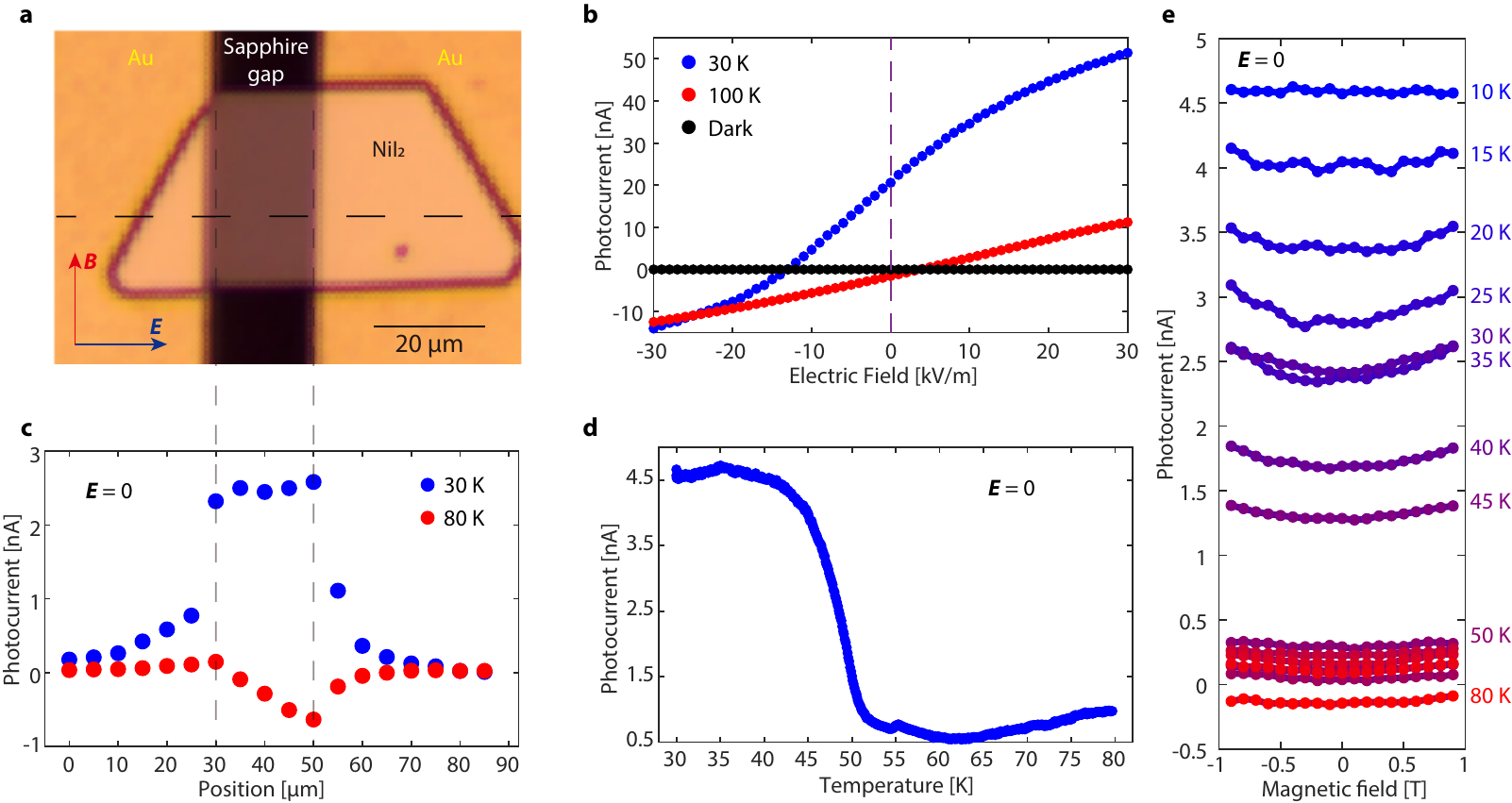}
    \caption{\textbf{Bulk photovoltaic effect (BPE) in bulk NiI$_2$}. \textbf{a}, Optical image of the PBE device. A PVD grown bulk-like NiI$_2$ flake was transferred across a sapphire gap, bridging two gold pads as electrodes. The electric field was applied between the electrodes and in a direction nearly parallel to the {\bf a}-axis, while the magnetic field was applied perpendicular to the electric field in plane. \textbf{b}, The electric field dependence of the photocurrent at 30 K, in the multiferroic phase and at 100 K, in the paramagnetic phase, reveals the presence of a polarization-induced internal electric field the multiferroic phase. \textbf{c}, The position dependence of the photocurrent along the dashed line in \textbf{a}, under zero bias shows a major contribution of the photocurrent from the NiI$_2$ between the electrodes. \textbf{d}, The temperature dependence of the zero-bias photocurrent shows a strong enhancement in the multiferroic phase. \textbf{e}, The external magnetic field increased the zero-bias photocurrent by 10-15\%, which we ascribe to an increase of the electric polarization from magnetoelectric coupling. Linearly polarized 532 nm light (0.3 mW power) was used in the BPE measurement.}
    \label{fig:figS6}
\end{figure}

\newpage

\begin{figure}[hbt!]
    \centering
    \renewcommand{\figurename}{Extended Data Figure}
    \includegraphics{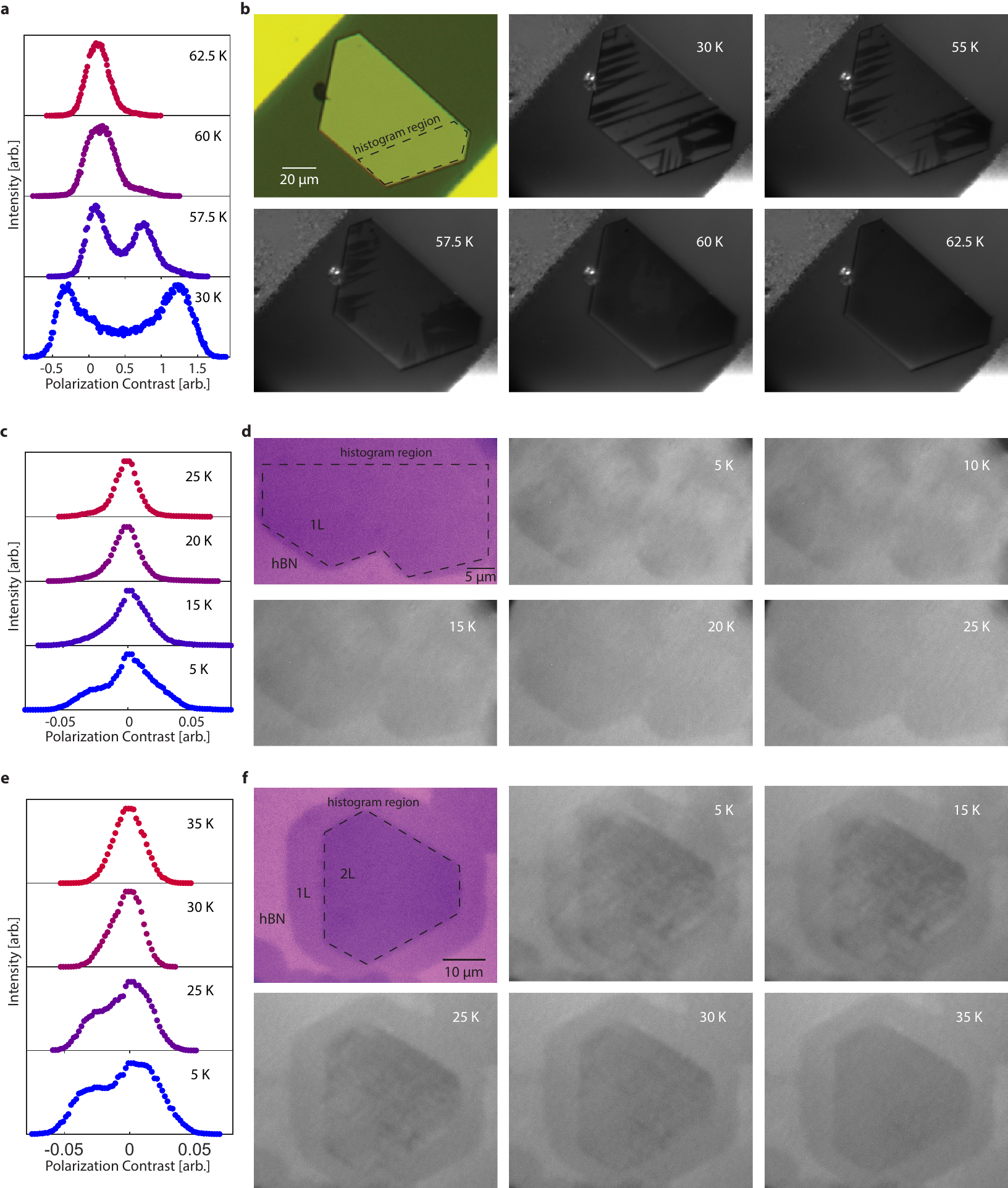}
    \caption{ \textbf{Cross-polarization images of the bulk, 1- and 2-layer NiI$_2$}. Temperature-dependent histogram plots of the polarization contrast images for \textbf{a} bulk, \textbf{c}, 1-layer region and \textbf{e}, 2-layer region. The upper-left images in \textbf{b, d, f} show optical images of bulk, 1- and 2-layer NiI$_2$ flakes, respectively. Subsequent images depict the temperature-dependent polarization contrast images for key temperatures. The raw cross-polarization images at various temperatures across the multiferroic transition provide signatures of the domain dynamics, explaining the spatial inhomogeneity of the transition temperature identified through polarization rotation measurements. The domain texture vanishes in \textbf{d}, 1-layer near 20 K, and in \textbf{f}, 2-layer near 35 K, consistent with polarization rotation measurements (Fig. \ref{fig:fig2}d).}
    \label{fig:figS7}
\end{figure}

\newpage

\begin{figure}[hbt!]
    \centering
    \renewcommand{\figurename}{Extended Data Figure}
    \includegraphics{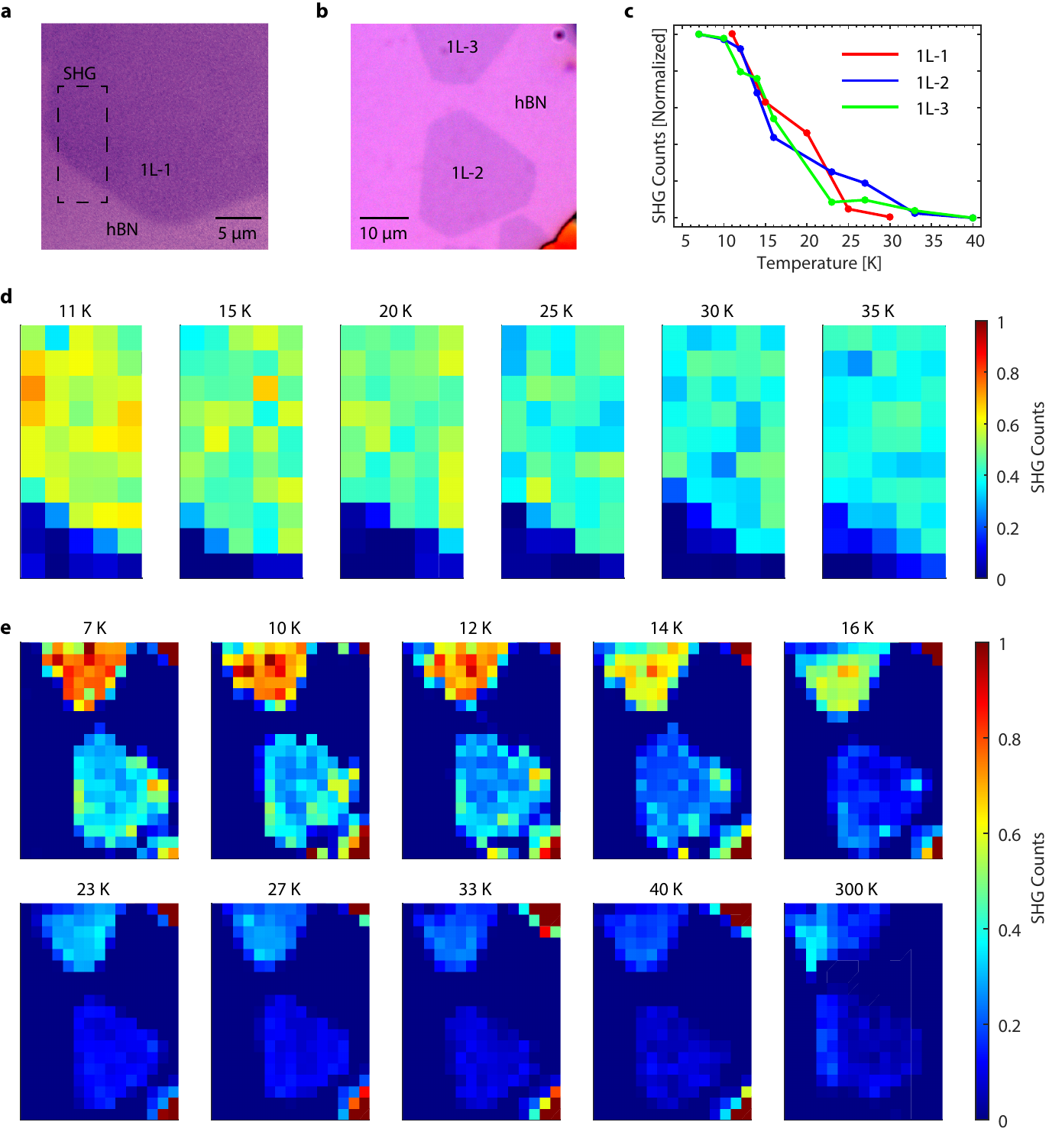}
    \caption{\textbf{Temperature dependent Second Harmonic Generation imaging of the single-layer NiI$_2$ crystals}. \textbf{a, b}, Optical images of the region where the SHG imaging was performed. \textbf{c}, Integrated SHG counts on three single-layer NiI$_2$ crystals show a transition around 20 K, which is consistent with the polarization rotation measurement. \textbf{d}, Temperature dependent SHG imaging of the rectangular region in \textbf{a} using 780 nm excitation. \textbf{e}, Temperature dependent SHG imaging of the region in \textbf{b} using 991 nm laser. Colorbar: SHG counts.}
    \label{fig:figS8}
\end{figure}

\newpage

\begin{figure}[hbt!]
    \centering
    \renewcommand{\figurename}{Extended Data Figure}
    \includegraphics{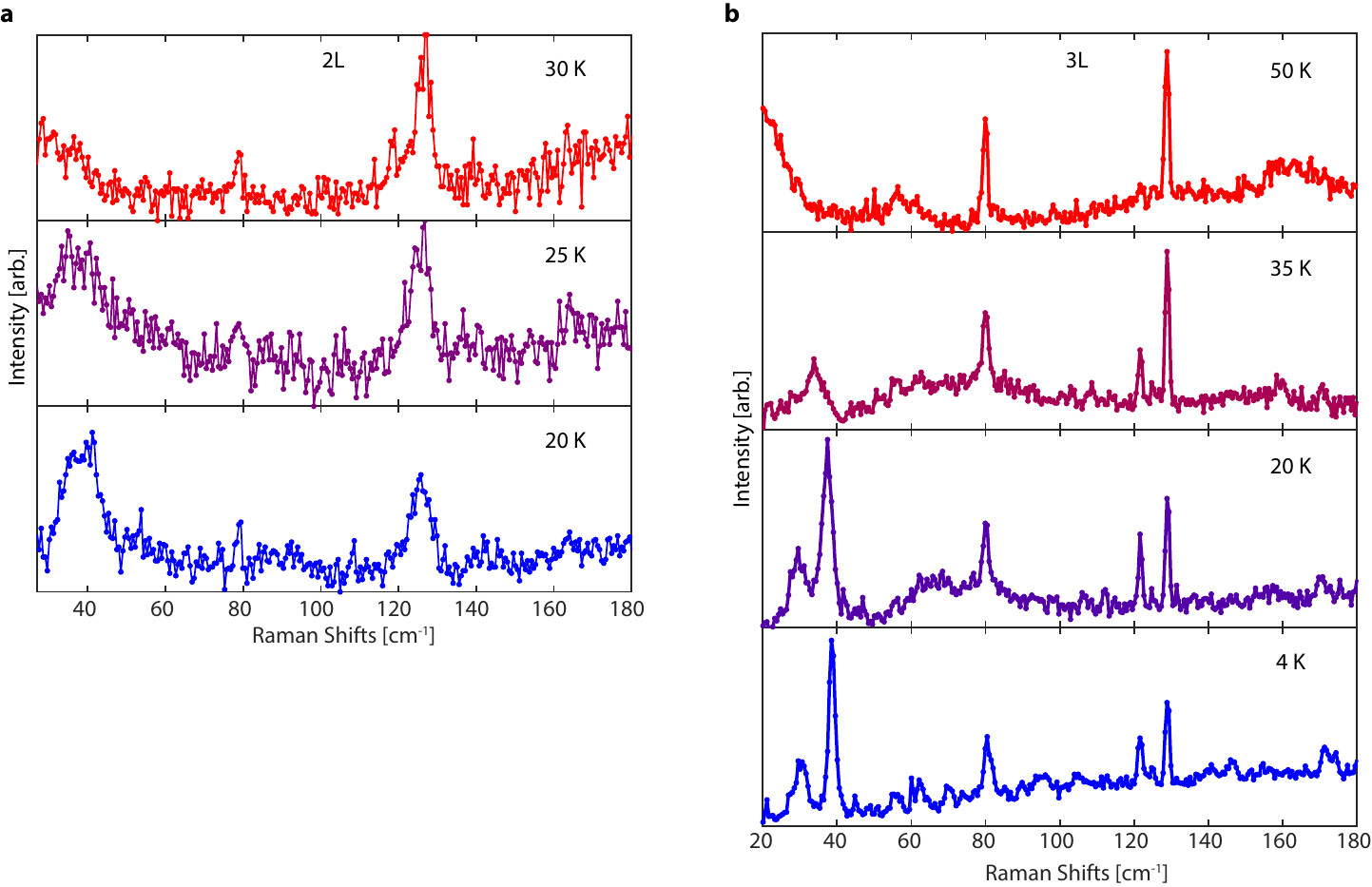}
    \caption{\textbf{Temperature dependent Raman Spectroscopy of two- and three-layer NiI$_2$ in cross-polarized (XY) configuration} The soft modes at around 38 cm$^{-1}$ in \textbf{a}, 2-layer and \textbf{b}, 3-layer built up below 25 K and 35 K respectively, which are consistent with the transition temperature measured from polarization rotation and SHG.}
    \label{fig:figS9}
\end{figure}

\newpage

\begin{figure}[hbt!]
    \centering
    \renewcommand{\figurename}{Extended Data Figure}
    \includegraphics{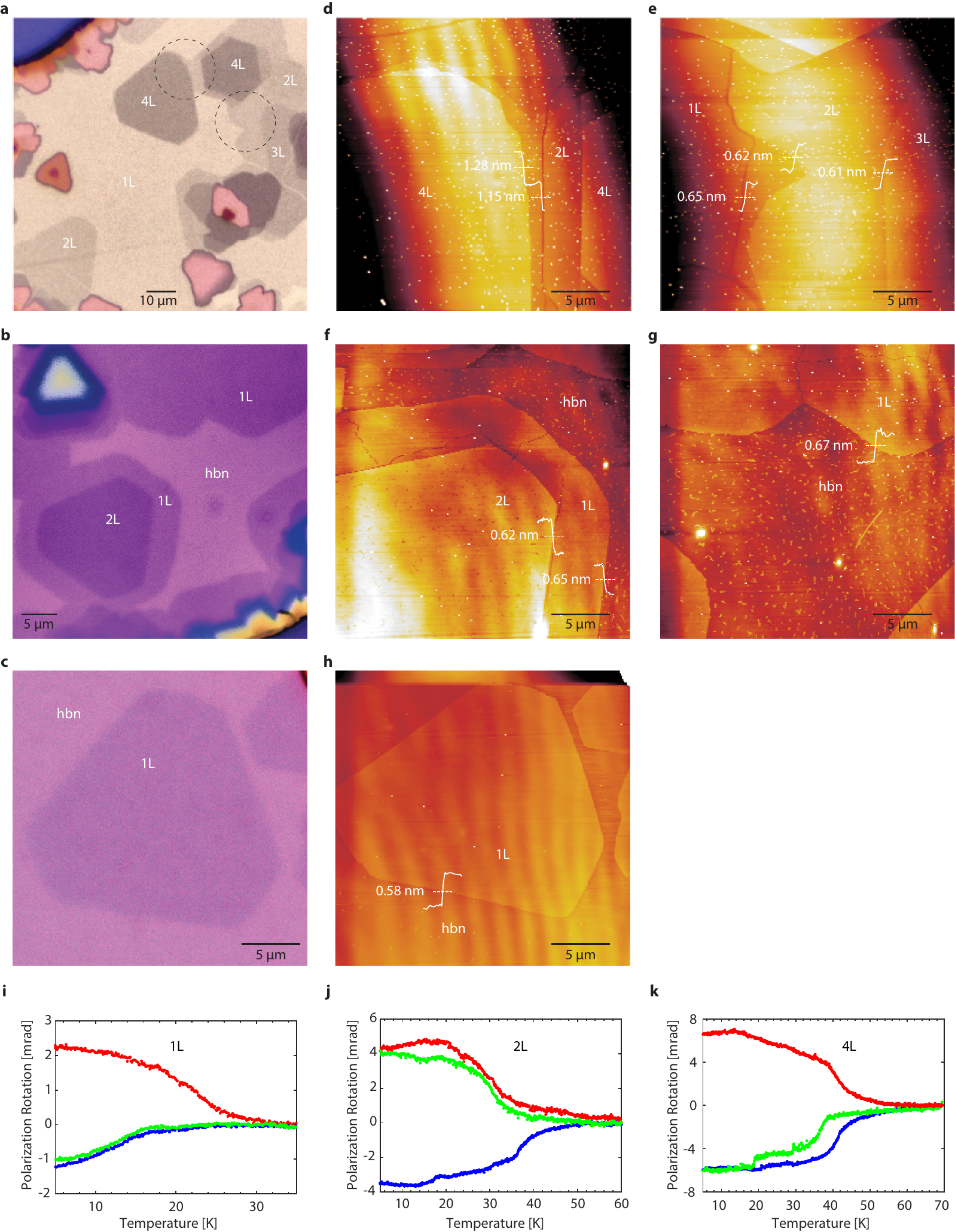}
    \caption{\textbf{Atomic Force Microscopy and additional polarization rotation measurements on few-layer NiI$_2$ crystals}. \textbf{a-c}, Wide-field optical images of one- to four-layer NiI$_2$ samples used in the optical measurements. \textbf{d-h}, Corresponding AFM maps of the region shown in the optical images. \textbf{b-f}, Temperature dependent polarization rotation measurements on one- to four-layer samples in different domain regions.}
    \label{fig:figS10}
\end{figure}

\end{document}